\documentclass[conference]{IEEEtran}
\IEEEoverridecommandlockouts
\usepackage{cite}
\usepackage{amsmath,amssymb,amsfonts}
\usepackage{algorithmic}
\usepackage{graphicx}
\usepackage{textcomp}
\usepackage{xcolor}
\def\BibTeX{{\rm B\kern-.05em{\sc i\kern-.025em b}\kern-.08em
    T\kern-.1667em\lower.7ex\hbox{E}\kern-.125emX}}
\usepackage{multirow}

\begin{document}

\title{\vspace{-12pt} \huge Enhancing Security of Memristor Computing System\\ 
\vspace{-2pt}
Through Secure Weight Mapping\\
\vspace{-17pt}
\thanks{This work was supported by the National Natural Science Foundation of China (No. 62172224) and the European Research Council through the European Union's Horizon 2020 Research and Innovation Programe under Grant 757259.}
}

\author{
	\IEEEauthorblockN{Minhui Zou\IEEEauthorrefmark{1}, Junlong Zhou\IEEEauthorrefmark{2}, Xiaotong Cui\IEEEauthorrefmark{3}, Wei Wang\IEEEauthorrefmark{1}, and Shahar Kvatinsky\IEEEauthorrefmark{1}}
    \IEEEauthorblockA{\IEEEauthorrefmark{2}School of Computer Science and Engineering, Nanjing University of Science and Technology\\\IEEEauthorrefmark{3}School of Cyber Security and Information Law, Chongqing University of Posts and Telecommunications\\\IEEEauthorrefmark{1}Faculty of Electrical and Computer Engineering, Technion - Israel Institute of Technology}
    \vspace{-25pt}
    }
\maketitle
\vspace{-30pt}

\IEEEpubid{\begin{minipage}{\textwidth}\ \\[12pt] \centering \copyright 2022 IEEE. Personal use of this material is permitted.  Permission from IEEE must be obtained for all other uses, in any current or future media, including reprinting/republishing this material for advertising or promotional purposes, creating new collective works, for resale or redistribution to servers or lists, or reuse of any copyrighted component of this work in other works. \end{minipage}} 

\vspace{-16pt}
\begin{abstract}
    Emerging memristor computing systems have demonstrated great promise in improving the energy efficiency of neural network (NN) algorithms.
	The NN weights stored in memristor crossbars, however, may face potential theft attacks due to the nonvolatility of the memristor devices.
	In this paper, we propose to protect the NN weights by mapping selected columns of them in the form of 1's complements and leaving the other columns in their original form, preventing the adversary from knowing the exact representation of each weight.
	The results show that compared with prior work, our method achieves effectiveness comparable to the best of them and reduces the hardware overhead by more than 18X.
\end{abstract}


\vspace{-3pt}
\section{Introduction}
\begin{figure}[ht]
    \centering
    \includegraphics[width=0.43\textwidth]{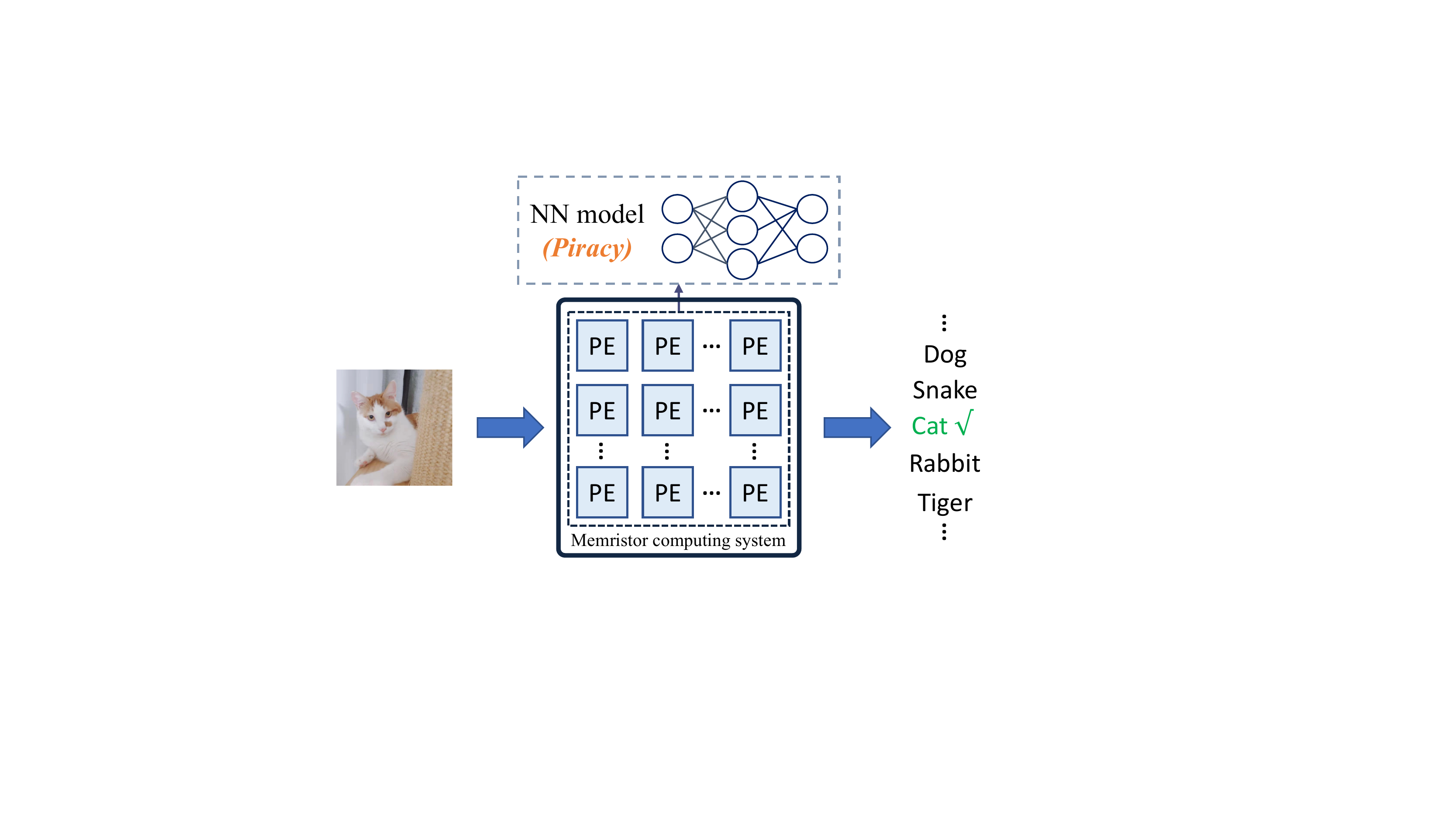}
    \vspace{-10pt}
    \caption{Structure of a memristor computing system and thwarting potential theft attacks by secure weight mapping.}
    \label{fig:Memory_Computing_System_basic_structure}
	\vspace{-22pt}
\end{figure}

\IEEEpubidadjcol
\vspace{-3pt}
Neural network (NN) algorithms are essential elements across industries such as robotics, visual object recognition, and natural language processing.
They are typically data-intensive, involving a large number of vector-matrix multiplications (VMMs).
Conventional computer architectures separating computation and memory are challenged because of their ineffectiveness in executing such algorithms.
The computing systems based on emerging memristor devices, such as resistive random-access memory (RRAM) and phase-change memory (PCM), introduce an in situ solution to improve speed and energy efficiency.
These memristor computing systems can both store the NN weights and process them in memory, which avoids a huge matrix data movement between computing units and memory \cite{wang_cross-point_2019}. 
As shown in Fig. \ref{fig:Memory_Computing_System_basic_structure}, a memristor computing system consists of many processing elements (PEs), each equipped with a group of memristive crossbar arrays and peripheral components.

The NN weights stored in the memristor crossbars, however, are exposed to the adversary when the system is turned off due to the nonvolatility of memristor devices.
With access to the memristor conductance values, the adversary may extract the well-trained NN models from them.
The extracted NN models could damage the intellectual property of the NN model designers and may lead to an information leak if the models are trained with private training datasets.
Even worse, trojans may be inserted into extracted NN models \cite{liu_neural_2017} and then implemented in critical areas such as facial recognition systems, which could cause security crises.
Therefore, it is important and urgent to protect the NN weights stored in memristor crossbars.

The straightforward solution is to encrypt the NN weights and decrypt them each time they are used. 
For example, \cite{li_p3m_2019} proposed to encrypt the whole NN model and \cite{cai_enabling_2019} proposed encrypting only part of the NN weights.
An incremental encryption method \cite{chhabra_i-nvmm_2011} could also be implemented since the NN inference is processed layer by layer.
The methods involve frequent writing operations to the memristor devices for encryption and decryption of NN weights.
Unfortunately, frequent writing operations to the memristor devices not only consume high energy and introduce long latency into the systems~\cite{chang_194_2014, yao_face_2017}, they also shorten the lifetime of the memristor computing systems due to the limited endurance of memristor devices~\cite{wang_cross-point_2019}.
\cite{zou_security_2020}, \cite{wang_low_2021}, and \cite{huang_new_2020} proposed to protect the NN weights through hiding either the row connections or column connections between memristor crossbars.
Such methods incur significant area and power overheads due to expensive implementation of the protection modules.
To tackle these problems, this paper proposes to enhance the security of memristor computing systems by defending against theft attacks aimed at the NN weights.
The contributions of this paper are summarized below:
\vspace{-4pt}
\begin{itemize}
  \item There are two popular schemes for mapping NN weights to memristor crossbars. 
  One is biasing the negative weights to be non-negative and the other is using the differential values of pairs of memristor devices to represent negative weights. For each  of the mapping schemes, we propose a protection method based on encoding selected columns of weights.
  \vspace{-2pt}
  \item We present the implementation of the two protection methods in memristor computing systems. 
  The proposed methods do not affect the performance of the systems. 
  We also suggest techniques to increase the security of the proposed methods by protecting every smaller block instead of every crossbar group and padding small weight matrices.
  \vspace{16pt}
  \item We compare the proposed methods with prior work on three NN models. 
  The experimental results show that the proposed methods achieve effectiveness on par with the best of the prior work and generate the lowest hardware overhead to the systems.
\end{itemize}

\vspace{-6pt}
\section{Background}
\label{sec:Threat_Model_Related_works_and_Motivation}
    \vspace{-4pt}
	\subsection{Threat Model and Motivation}
	    \vspace{-4pt}
		We assume that the adversary knows the structures of the NN models stored in the memristor computing systems. 
		We also assume that the adversary could access the conductance values of the memristor devices \cite{huang_new_2020}. 
		With both the NN structures and weights, the adversary could extract the well-trained NN models.
		Hence, in this paper, the goal of the adversary is to read the NN weights correctly from the memristor devices.
		We, in opposition, want to prevent the adversary from realizing this goal.
		As shown in Fig. \ref{fig:Memory_Computing_System_basic_structure}, the NN weights are securely mapped to crossbars so that the NN models extracted by the adversary could not function normally.
	
	\vspace{-6pt}	
	\subsection{Related Work}
		\cite{zou_security_2020} proposed to obfuscate the crossbar row connections between the positive and negative crossbars.
		\cite{wang_low_2021} suggested hiding the crossbar column connections between consecutive crossbars.
		Both these protection methods, however, rely on high-multiplicity multipliers and demultipliers, which generate a large hardware overhead.
		\cite{huang_new_2020} proposed to hide the connections between the inputs and the crossbar rows.
		This protection method is based on SRAM arrays, which demand a significant amount of hardware area and power.
		In Section \ref{sec:Experiments}, we show that our proposed methods achieve  effectiveness comparable to the best of the related work and are more efficient than them in terms of area and power.

    \vspace{-6pt}
	\subsection{Preliminaries}
		The most computing-intensive and time-consuming parts of NN alogrithms are convolution (Conv) layers and fully-connected (FC) layers.
		The main computation of FC layers can be implemented directly with VMMs, which are described as: 
		\vspace{-6pt}
		\begin{equation} \label{eq:fc_operations}
		    \vspace{-6pt}
			y_j = \sum_{i=0}^{m-1} w_{i,j} \cdot x_{i},
		\end{equation}
		where $x_{i}(i \in [0,m-1])$ is the input feature map, $y_j(j \in [0,n-1])$ is the output, and $w_{i,j}$ is the synapse weight.
		The main computation of Conv layers is different but could also be transformed to be implemented with VMMs.
		
		In memristor computing systems, the input feature maps are transformed into voltages $(\textit{\textbf{V}})$ by using digital-to-analog converters (DACs) that are applied to the wordlines (WLs) of the memristor crossbars.
		The bitlines (BLs) of the memristor crossbars output the accumulated currents $(\textit{\textbf{I}})$.
		The output currents are then transformed by using analog-to-digital converters (ADCs).
		The analog VMMs performed by a memristor crossbar are described as: 
        \vspace{-6pt}
		\begin{equation} \label{eq:analog_VMM}
			I_j = \sum_{i=0}^{m-1} c_{i,j} \cdot V_{i},
			\vspace{-6pt}
		\end{equation}
		where $c_{i,j}$ is the conductance of the cell at the $i$th row and $j$th column of the crossbar.
		Due to the nonlinear imperfection and immature manufacturing of memristor devices, the conductance value of a memristor device can only be tuned to limited discrete conductance states, and multiple crossbars are used to represent high-precision weights \cite{zhu_configurable_2019}.

\section{NN weight mapping schemes and protection methods}
	An NN weight can be positive or negative, but the conductance of memristor devices can only be positive.
	To support negative weights, different mapping schemes were proposed \cite{shafiee_isaac_2016,zhu_configurable_2019}.
	Two popular mapping schemes are biasing the negative weights to be non-negative \cite{shafiee_isaac_2016} and using the differential values of pairs of memristor devices to represent negative weights \cite{zhu_configurable_2019}.
	For the purposes of our discussion, let us denote the two mapping schemes as mapping scheme 1 and mapping scheme 2, respectively.
	Based on these two mapping schemes, we present two corresponding NN weight protection methods.
    
    \vspace{-6pt}
	\subsection{Mapping Scheme 1 and Proposed Protection}
		\begin{figure*}[ht]
			\centering
			\includegraphics[width=0.68\textwidth]{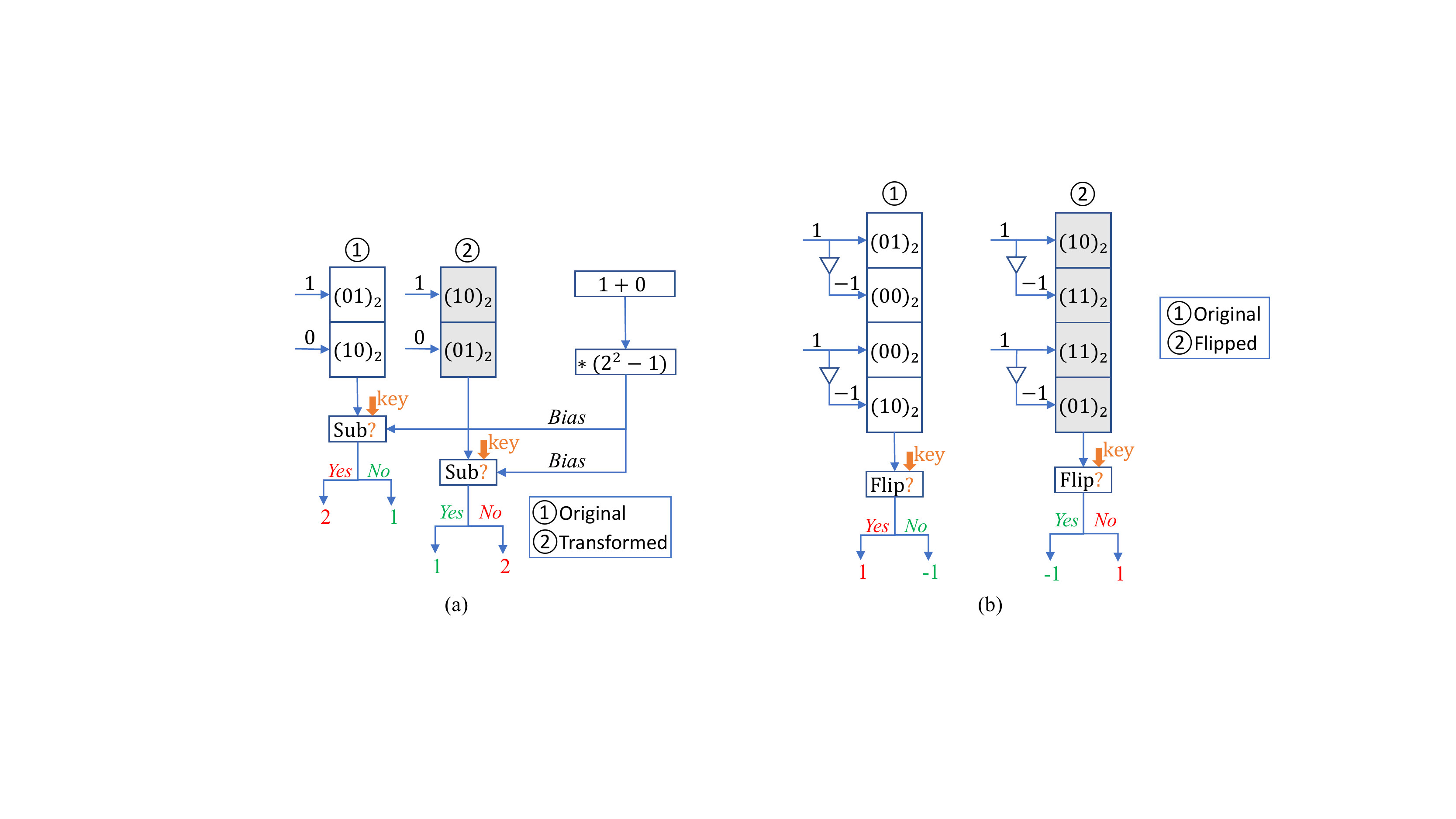}
			\vspace{-10pt}
			\caption{Two simple examples explaining (a) the protection method for mapping scheme 1, and (b) the protection method for mapping scheme 2.}
			\label{fig:simple_examples}
			\vspace{-18pt}
		\end{figure*}
		
		\cite{shafiee_isaac_2016} proposed to add a bias to all the NN weights so that they are all non-negative.
		After computation, the bias is then subtracted to restore the correct results.
		For this mapping scheme, we propose to protect the NN weights by selectively encoding the weights. 
    	Fig. \ref{fig:simple_examples}(a) shows an example, where an input vector $(1, 0)$ is multiplying a column of weights $({01}_2,{10}_2)^T$, which is mapped to column \raisebox{.5pt}{\textcircled{\raisebox{-.9pt} {1}}} and column \raisebox{.5pt}{\textcircled{\raisebox{-.9pt} {2}}}, of memristor devices, respectively.
    	Column \raisebox{.5pt}{\textcircled{\raisebox{-.9pt} {1}}} presents the weights mapped in the original form.
		Column \raisebox{.5pt}{\textcircled{\raisebox{-.9pt} {2}}} presents the weights mapped in the form of 1's complements.
		The 1's complement of weight $w_{i,j}$ is determined by
		\vspace{-6pt}
		\begin{equation} \label{eq:2s_complement_wij}
			\hat w_{i,j} = 2^{p_w} -1 - w_{i,j},
			\vspace{-6pt}
		\end{equation}
		where $p_w$ denotes the precision of the weight $w_{i,j}$.
		We denote the weights in the form of 1's complements as transformed weights.
		With the transformed weights, the observed VMM output for the $j$th column of weights could be written as
		\vspace{-6pt}
		\begin{equation} \label{eq:fc_operations_2s_complement}
			\hat y_j = \sum_{i=0}^{m-1} \hat w_{i,j} \cdot x_{i}.
		\vspace{-6pt}
		\end{equation}
		Combining (\ref{eq:fc_operations}), (\ref{eq:2s_complement_wij}) and (\ref{eq:fc_operations_2s_complement}), we get
		\vspace{-6pt}
		\begin{equation} \label{eq:fc_operations_complementory_get_yj}
			y_j = (2^{p_w}-1)\sum_{i=0}^{m-1} x_{i} - \hat y_j.
			\vspace{-6pt}
		\end{equation}
		Note that $\sum_{i=0}^{m-1} x_{i}$ is the sum of inputs.
		For this example, the sum of inputs is $1+0$.
    	Also note that $(2^{p_w}-1)$ is a constant scalar, and in this example, the scalar is $2^2-1$.
    	The product of the sum of the inputs and the scalar is the bias, which subtracts the observed column output $\hat y_j$ to restore the correct VMM result.
    	Column \raisebox{.5pt}{\textcircled{\raisebox{-.9pt} {1}}} and column \raisebox{.5pt}{\textcircled{\raisebox{-.9pt} {2}}} of the weights receives the same input vector.
    	For column \raisebox{.5pt}{\textcircled{\raisebox{-.9pt} {1}}} of weights, the correct VMM result is just the observed output, while for column \raisebox{.5pt}{\textcircled{\raisebox{-.9pt} {2}}} the correct VMM result is calculated by subtracting the observed output from the bias.
    	The adversary, however, does not know which column of weights are mapped in the transformed way.
    	Thus, the adversary has to guess the key for each column.
    	There are two options for each column of memristor devices.
    	The correct keys are $No$ and $Yes$ for column \raisebox{.5pt}{\textcircled{\raisebox{-.9pt} {1}}} and column \raisebox{.5pt}{\textcircled{\raisebox{-.9pt} {2}}}, respectively.
    	If the input keys are not correct, the VMM result will be wrong.

	\subsection{Mapping Scheme 2 and Proposed Protection}
		An NN weight $w_{i,j}$ could also be implemented with the differential value of a pair of memristor devices \cite{zhu_configurable_2019}, one of which is connected with a positive voltage and the other of which a negative voltage.
		We denote the conductance of the two memristor devices as $g^+_{i,j}$, and $g^-_{i,j}$, respectively.
		Now (\ref{eq:fc_operations}) is rewritten as
		\vspace{-10pt}
		\begin{equation} \label{eq:fc_operations_pos_neg}
			y_j = \sum_{i=0}^{m-1} (c^+_{i,j}-c^-_{i,j}) \cdot x_{i}.
			\vspace{-6pt}
		\end{equation}

		Regarding this mapping scheme, we propose to protect the NN weight by encoding the memristor conductance instead of the weights.
		That is 
		\vspace{-6pt}
		\begin{equation} \label{eq:pos_neg_ones_complement}
			\hat c^{\pm}_{i,j} = 2^b - 1 - c^{\pm}_{i,j}.
			\vspace{-6pt}
		\end{equation}
		Fig. \ref{fig:simple_examples}(b) shows an example, where an input vector $(1, 1)$ is multiplying a column of weights $({01}_2,-{10}_2)^T$, mapped to pair \raisebox{.5pt}{\textcircled{\raisebox{-.9pt} {1}}} and pair \raisebox{.5pt}{\textcircled{\raisebox{-.9pt} {2}}} of memristor columns, respectively.
    	The pair \raisebox{.5pt}{\textcircled{\raisebox{-.9pt} {1}}} is the weights mapped in the memristor devices in the original form.
		The pair \raisebox{.5pt}{\textcircled{\raisebox{-.9pt} {2}}} is in the form of 1's complement, the transformed memristor conductance of pair \raisebox{.5pt}{\textcircled{\raisebox{-.9pt} {1}}}.
		Combining (\ref{eq:fc_operations_pos_neg}) and (\ref{eq:pos_neg_ones_complement}), we get
		\vspace{-6pt}
		\begin{equation} \label{eq:fc_operations_pos_neg_ones_complement_get_yi}
			y_j = - \hat y_j, 
			\vspace{-6pt}
		\end{equation}
		where we can see that when the memristor conductance of a pair of columns is encoded, the VMM result is simply the reverse of the original VMM result.
		Thus, in order to get the correct VMM result of the transformed column of weights, we just need to flip the sign bit of the observed output.
		For the pair of columns \raisebox{.5pt}{\textcircled{\raisebox{-.9pt} {1}}}, the correct VMM result is the observed output without being flipped, while for pair of columns \raisebox{.5pt}{\textcircled{\raisebox{-.9pt} {2}}}, it is flipped.
    	Nevertheless, again, the adversary only has to guess which column of weights are mapped in the transformed way.
    	Likewise, there are also two options for each pair of memristor columns and the VMM result will be correct only when the input keys are correct.

\section{Implementing the Protection Methods}
    We follow the MNSIM~\cite{zhu_mnsim_2020} architecture.
	A PE contains $G$ crossbar groups and each such group consists of a single crossbar for mapping scheme 1 or a crossbar pair for mapping scheme 2.
	Furthermore, each crossbar group is equipped with some peripheral components such as ADCs and Shfit\&Adds.
	The proposed methods map selected columns of NN weights in the transformed form, leaving the other columns in their original form.
	
	\subsection{Mapping Scheme 1}
		\begin{figure*}[ht]
			\centering
			\includegraphics[width=0.82\textwidth]{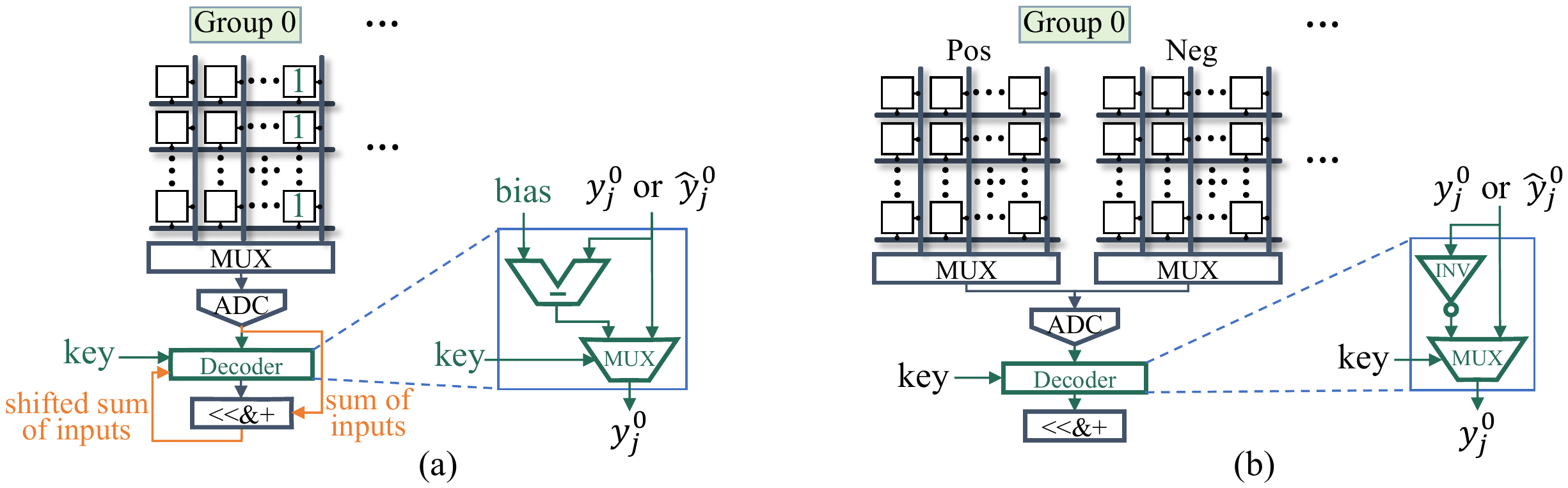}
			\vspace{-10pt}
			\caption{Design of the proposed protection methods: (a) For mapping scheme 1, designing and inserting the decoder module for each crossbar group with a single crossbar; and (b) for the mapping scheme 2, designing and inserting the decoder module for each crossbar group with a positive/negative crossbar pair.}
			\label{fig:implementation_of_methods}
			\vspace{-17pt}
		\end{figure*}

		To correctly output the VMM results of the columns of weights mapped in the transformed form, a decoder module is inserted in the PE and is connected to the ADC, as shown in Fig. \ref{fig:implementation_of_methods}(a).
		The decoder module comprises a full subtractor and a 2:1 MUX.
		
		Assume a column of weights $w_{:,j}$ with its size equal to the height of the memristor crossbar.
		Assume also that the precision of the memristor devices is $p_m$ and $p_w=p_m \times G$.
		Then $w_{:,j}$ is divided into $g$ columns of partial weights ($w^0_{:,j}$ to $w^{G-1}_{:,j}$) and these are mapped to the $j$th memristor column across the $G$ crossbar groups, respectively.
		The VMM result for $w^g_{:,j}$ ($g \in [0,G-1]$) is 
		
		\vspace{-8pt}
		\begin{equation} \label{eq:divide_yj}
			y_j = \sum_{g=0}^{G-1} { {(2^{p_m})}^{G-1-g} \cdot {y^g_j} },
			\vspace{-6pt}
		\end{equation}
		where $y^g_j$ is the partial VMM result of the $j$th crossbar column in the $g$th crossbar group.
		If $w_{:,j}$ is mapped in the transformed form, then the partial VMM result for $w^g_{:,j}$ is 
        \vspace{-6pt}
		\begin{equation} \label{eq:divide_yj_transformed}
			y^g_j = (2^{p_m}-1) \sum_{i=0}^{m-1} x_{i} - \sum_{i=0}^{m-1} {\hat w^g_{i,j}} \cdot x_{i},
			\vspace{-6pt}
		\end{equation}
		where ${\hat w^g_{i,j}}$ is the 1's complement of $w^g_{i,j}$, and
		$\sum_{i=0}^{m-1} x_{i}$ is the sum of inputs, which can be calculated by the inputs multiplied by a column of memristor devices whose values are all being fixed as 1, as shown in Fig. \ref{fig:implementation_of_methods}(a).
		$2^{p_m} \sum_{i=0}^{m-1} x_{i}$ can be calculated by shifting the sum of inputs left by ${p_m}$ bits using the Shift\&Add inside the PE.
		The bias $(2^{p_m}-1) \sum_{i=0}^{m-1} x_{i}$ can then be calculated by subtracting the sum of inputs from the shifted sum of inputs by using the subtractor of the decoder module, as in Fig. \ref{fig:implementation_of_methods}(a).
		The bias is stored in a register or buffer within the PE and sent to the decoder module to restore the correct partial VMM result $y^g_j$.
		The bias only needs to be calculated once for each input vector.
		Note the key for the column of weights $w_{:,j}$ is shared by all the columns of partial weights ($w^0_{:,j}$ to $w^{G-1}_{:,j}$).

		Also note that for mapping scheme 1, all the NN weights are biased.
		To get the correct layer outputs of an NN layer, the system needs to collect the sum of the layer inputs to restore the unbiased layer outputs.
		\cite{shafiee_isaac_2016} proposed using the last memristor column of each crossbar to calculate the sum of inputs.
		Thus we could reuse the sum of inputs for the proposed method so that no extra memristor devices are required.

		A column of NN weights larger than the number of rows of a crossbar will be divided into multiple PEs, each generating a partial VMM output.
		Only when each partial VMM ouput is decoded correctly, will the column output be correct.
		For a column of NN weights with its size less than the number of rows of a crossbar, we pad it as further explained in Section \ref{sec:Increasing_the_security_of_the_proposed_protection_methods}.

		In order not to affect the system throughput, we insert one decoder module for each ADC.
		Because of the high overhead of ADCs and their relatively higher operation frequency than memristor reading, there are only a limited number of ADCs for each crossbar group, which are shared by the crossbar BLs throught multipliers \cite{wang_integration_2020}.
		Thus, for each crossbar group, the number of decoder modules, equal to the number of ADCs of the crossbar group, is small.
		For each column across all the $G$ crossbar groups, we need a 1-bit key.
		Each column output is checked to determine whether we have to do the subtraction according to the key.
	
	\vspace{-6pt}
	\subsection{Mapping Scheme 2}
	    \vspace{-4pt}
		For mapping scheme 2, each crossbar group contains a positive/negative crossbar pair.
		The outputs from the positive and negative crossbar pair are summed in the analog domain.
		Similar to the design of the protection method for mapping scheme 1, a decoder module is inserted in to the PE and connected to the ADC, as shown in Fig. \ref{fig:implementation_of_methods}(b).
		The decoder module comprises an inverter and a 2:1 multiplier.

		Similarly, a column of weights $w_{:,j}$ are divided into $G$ columns of partial weights ($w^0_{:,j}$ to $w^{G-1}_{:,j}$) and each of them is mapped to the $j$th positive/negative column pair of memristors of all the $G$ crossbar groups.
		For example, the column of partial weights $w^g_{:,j}$ ($g \in [0,G-1]$) is mapped to a positive column of memristors and a negative column of memristors, the conductance of which are denoted as $c^{g+}_{:,j}$, and $c^{g-}_{:,j}$, respectively.
		If the column of weights $w_{:,j}$ is selected for transformation, then both $c^{g+}_{:,j}$, and $c^{g-}_{:,j}$ are encoded, respectively. 
		The correct partial VMM output for $w^g_{:,j}$ is  
		\vspace{-8pt}
		\begin{equation} \label{eq:divide_yj_transformed__}
			y^g_j = - \sum_{i=0}^{m-1} ( {\hat c^{g+}_{i,j}} -  {\hat c^{g-}_{i,j}} ) \cdot x_{i},
			\vspace{-6pt}
		\end{equation}
		where ${\hat c^{g+}_{i,j}}$ and $ {\hat c^{g-}_{i,j}}$ are the 1's complements of ${c^{g+}_{i,j}}$ and ${c^{g-}_{i,j}}$, respectively.
		That is, the decode module needs to reverse the observed output to get ${y^g_j}$.
		Again, the key for the column of weights $w_{:,j}$ is 1-bit and shared by all the columns of partial weights ($w^0_{:,j}$ to $w^{G-1}_{:,j}$).
		The number of decoder modules inserted for each crossbar group is equal to the number of ADCs it has and the throughput of the ADCs is not affected.

    \vspace{-6pt}
	\subsection{Increasing the Security of the Proposed Protection Methods}
	\vspace{-4pt}
	    \label{sec:Increasing_the_security_of_the_proposed_protection_methods}
		Assume the memristor crossbars are $M \times N$ in size.
		For the protection methods for mapping schemes 1 and 2, the trial times for brutal forcing a single crossbar group is $2^{(N-1)}$ and $2^N$, respectively.
		To increase the security of the proposed methods, we propose to divide a crossbar (pair) into $k$ blocks by every $x$ WLs ($k=M/x$). 
		In fact, due to the resolution limitation of ADCs, each time only part of the WLs are activated \cite{chen_65nm_2018}.
		We set $x$ as the integer multiple of the number of activated WLs each time so that the crossbar performance is not affected.
		The security of the protection methods for mapping schemes 1 and 2 becomes $2^{k \times (N-1)}$ and $2^{k \times N}$, respectively.
		For example, if $M$, $N$, and $x$ are 128, 128, and 8, respectively, then the adversary needs to try $2^{1016}$ and $2^{1024}$ times to break a crossbar (pair) for the respective protection methods for mapping schemes 1 and 2, which is sufficiently secure for current computing power.

		For an NN weight matrix that is smaller than the size of a memristor crossbar, the security is limited.
		For example, again if $M$, $N$, and $x$ are 128, 128, and 8, respectively, the trial times to break a weight matrix of size $32 \times 32$ is only $2^{32\times(32/8)}$.
		To increase the security of a small NN weight matrix, we could pad it with random values within the range of the real minimal and maximal weight values so that it is the same size as the crossbars.
		The fake weight rows (columns) are permuted with the real weight rows (columns).
		The relative order of the real rows (columns), however, remains unchanged from that of the original small weight matrix.  
		An additional $M$-bit plus $N$-bit keys are needed to indicate the real rows and columns of weights, respectively.
	
	\vspace{-8pt}
	\subsection{Impact on the System Performance}
	\vspace{-4pt}
	For a given NN model, the NN weights are securely mapped to the memristor computing system based on the proposed protection methods according to the predefined keys.
	The mapping is done offline and only once.
	We assume the keys are stored in a tamper-proof memory (TPM) embedded in the system.
	After setup, each time the memristor computing system is used for inference, the decoder modules are provided with the keys from the TPM.
    The inserted decoder modules connected with the ADCs are non-blocking and the throughput of the ADCs is not affected. 
    For each decoder module for mapping scheme 1, only a single cycle is needed to calculate the bias for each crossbar input vector. 
    The additional 1-cycle latency could easily be hidden by system parallelism. 
    The decoder modules perform the subtraction between the bias and the column outputs or just pass through the column outputs, which does not incur any latency overhead. 
    For each decoder module for mapping scheme 2, it reverses or just passes through each column output, which does not incur any latency overhead, either.  
	\vspace{-8pt}
	\subsection{Security Analysis}
	\vspace{-4pt}
	One potential attack is dividing and conquering. 
	That is, each time the adversary targets only a single crossbar (group) column and expects to see higher and lower classification accuracy of the systems with the correct and wrong keys for the column, respectively. 
	In this way, the key for the column is discovered. 
	Then the adversary targets the next column and continues thus until the keys for all columns are discovered. 
	This kind of attack, however, does not work in our proposed technique, because when the majority of the NN weights of the whole NN models are protected, the classification accuracy of the systems stays around $10\%$ (for the CIFAR10 dataset \cite{krizhevsky_learning_2009}). 
	In fact, through experiments, we found removing the protection for a number of columns or a whole NN layer does not necessarily increase the classifying accuracy of the systems.
	It may even slightly decrease the classifying accuracy because of the non-linear characteristics of the NN algorithms. 
    To conclude, our proposed protection methods are immune from the divide-and-conquer type of attack. 

\vspace{-12pt}	
\section{Evaluation}
\IEEEpubidadjcol
\label{sec:Experiments}
    \vspace{-6pt}
	We tested the proposed protection methods on three NN models: LeNet\cite{lecun_gradient-based_1998}, AlexNet\cite{krizhevsky_imagenet_2012}, and VGG16\cite{simonyan_very_2015}. 
	These models are modified and trained on the CIFAR10 dataset with 8-bit weights.
	The accuracy of LeNet, AlexNet, and VGG16 is 70.27\%, 87.01\%, and 93.09\%, respectively.
	For comparison, we implemented the protection methods of \cite{zou_security_2020}, \cite{wang_low_2021}, and \cite{huang_new_2020} on the same models.
	The models are simulated with MNSIM\cite{zhu_mnsim_2020}. 
	The crossbar size for mapping scheme 1 is $256 \times 256$ and for mapping scheme 2 is $256 \times 257$ with the last column preserved for calculating the sum of inputs.
	Each PE has 8 crossbar groups and each crossbar group is equipped with 16 ADCs.
	Each time 16 WLs are activated.
	$x$ is set as 32.
	The precision of memristor devices is 1 bit.
	The area of memristor devices we used for simulation refers to \cite{wu_suppress_2018}, and the digital part of the system together with the protection modules of all the protection methods are evaluated based on 28nm CMOS technology.

	
	The protection method of \cite{zou_security_2020} only applies to crossbar pairs with crossbar row connections.
	Hence, the protection method of \cite{zou_security_2020} was only evaluated for mapping scheme 2.
	The other two protection methods \cite{wang_low_2021}, \cite{huang_new_2020} were implemented for both mapping schemes.
	For simplicity, we denote the mapping schemes 1 and 2 as m1 and m2, respectively; and denote the protection methods of the proposed, \cite{zou_security_2020}, \cite{wang_low_2021}, and \cite{huang_new_2020} as our, date20, asp21, and sram20, respectively.

    \vspace{-12pt}
	\subsection{Effectiveness}
\begin{figure}[t]
			\centering
			\includegraphics[width=0.45\textwidth]{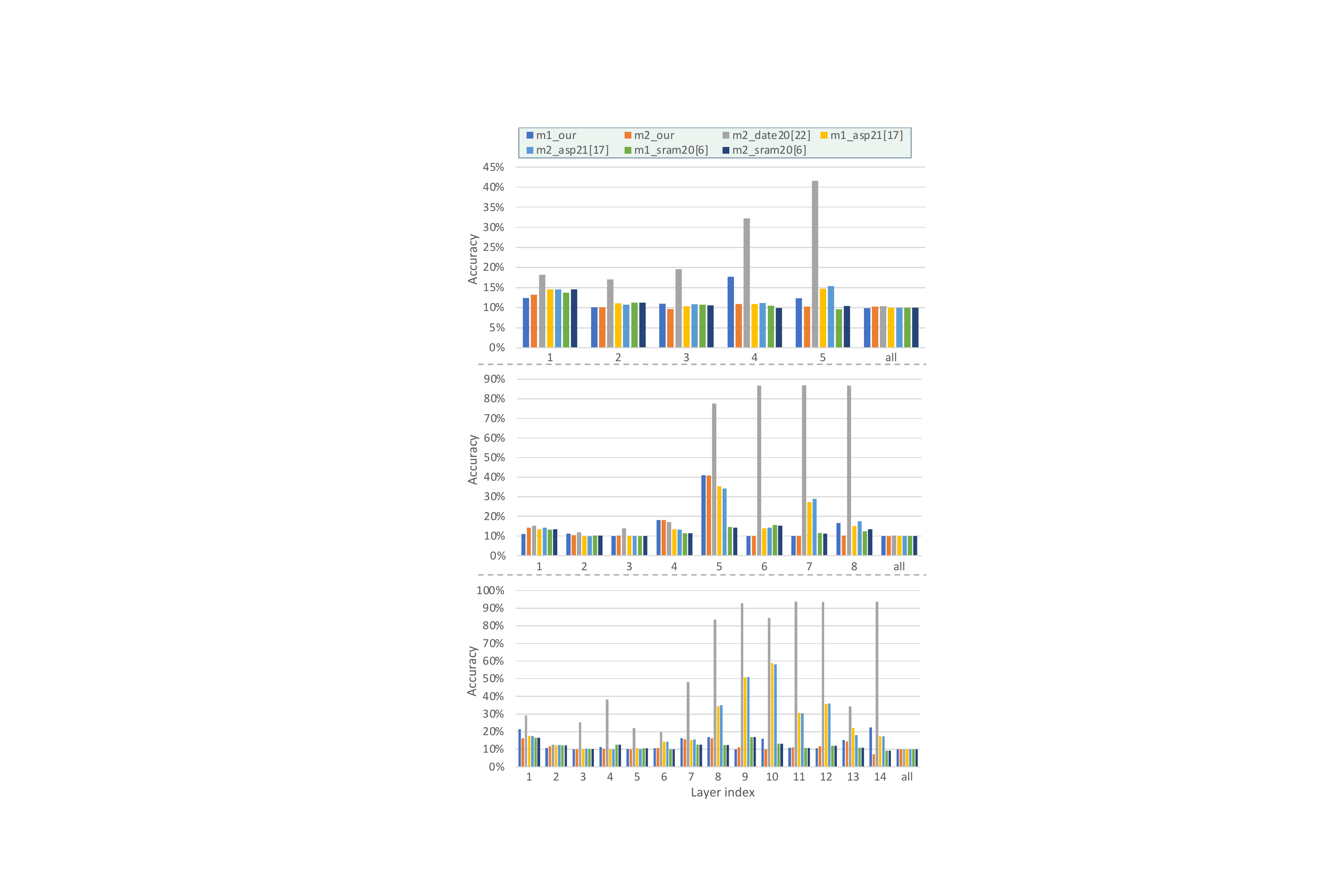}
			\vspace{-10pt}
			\caption{Results comparison of the effectiveness of the protection methods (lower is better) when protecting only a single layer or all layers for \textbf{\textit{Top}}: LeNet, \textbf{\textit{Middle}}: AlexNet, and \textbf{\textit{Bottom}}: VGG16.}
			\label{fig:result_layer_significance}
			\vspace{-18pt}
		\end{figure}

	    \vspace{-4pt}
		The effectiveness of the proposed methods is defined as the prediction accuracy of the extracted NN models by using random keys.
		Each experiment was carried out 40 times and the average results are determined.
		The lower the accuracy is, the better the effectiveness of the method. 
		Fig. \ref{fig:result_layer_significance} shows the effectiveness of all the protection methods by protecting only a single NN layer or all NN layers.
		When all the layers are protected, all the protection methods could lead to the accuracy of the NN models dropping to approximately 10\%, which means the NN models are no better than random guessing.
		On the other hand, when only a single layer is protected, the effectiveness of the protection methods varies.
		For all the models, when only protecting a single layer, \cite{zou_security_2020} could not protect them well for some layers such as layer 4 of LeNet while the other methods remain effective for all layers.
		For small layers or layers of small models, the proposed methods together with \cite{wang_low_2021} and \cite{huang_new_2020} remain effective.
		For the large layers of the big model VGG16, the proposed methods outperform \cite{wang_low_2021}.
		Overall, the effectiveness of the proposed methods is comparable with that of the most effective method \cite{huang_new_2020}.

    \vspace{-8pt}
	\subsection{Hardware Overhead}
	    \vspace{-4pt}
		The hardware overhead of the protection methods was evaluated in terms of the area, power, and key storage of the protection modules inserted in the memristor computing systems.
		For simplicity, assume the bitwidth of VMM inputs and outputs are both 8 bits.
		For the proposed methods, the protection modules are the decoder modules and the number of required protection modules for each crossbar group, which equals that of the ADCs of each crossbar group.
		For \cite{zou_security_2020}, the protection module comprises 32 16:1 MUXes and 16 1:16 DEMUXes and each crossbar group needs one protection module.
		For each 16 WLs, the key storage for each protection module is $48 \times 4$ bits (each MUX or DEMUX needs 4 bits) and so the key storage for each protection module is $48\times 4\times 16$ bits.
		For \cite{wang_low_2021}, the size of a virtual operating unit (VOU) is scaled as $16 \times 16$.
		The protection module includes one 16:1 MUX and one 1:16 DEMUX and each crossbar group needs one protection module.
		The authors, however, did not consider the module's bitwidth in their paper.
		In fact, a single input (output) or output (single input) of the MUX (DEMUX) is an array of 16 8-bit values. 
		For fairness of comparison, we set the bitwidth of the MUX and DEMUX in their method as $16\times 8$.
		The key storage for each protection module is ($256\times 8 + 4\times 2\times 16$) bits (row activation vectors and the keys for MUX/DEMUX for each 16 VOUs).
		For \cite{huang_new_2020}, the protection module is a $256 \times 256$ SRAM array and each crossbar group needs one protection module.
		The key storage for each protection module is $256 \times 8$ bits (the locations of the "1" bits).
		We assume, for the methods of \cite{zou_security_2020,wang_low_2021,huang_new_2020}, the keys are shared among all the protection modules inside each PE to reduce key storage.

		Table \ref{tab:overhead_area_power} lists the area and power overheads of our proposed methods.
		For example, the first cell means the area overhead of the proposed method for mapping scheme 1 for the LeNet model is 0.0048\%.
		The area and power overheads of the proposed methods are less than 0.12\% and 0.10\% for all models, respectively.
		The area and power overheads of the proposed method for mapping scheme 2 are less than that of the proposed method for mapping scheme 1.
		The reason is that for mapping scheme 1, each crossbar group only has one crossbar while for mapping scheme 2, each crossbar group has one pair of crossbars.
		Table \ref{tab:overhead_comparison_normalized} lists the normalized result comparison of area, power, and key storage overhead for the proposed methods and the related work.
		The proposed methods cost much less in terms of area and power overheads compared with the other protection methods.
		The reason is the protection modules of the proposed methods are made up of simpler subtractors, inverters and low-multiplicity MUXes.
		\cite{huang_new_2020} incurs the largest area and power overheads because it is based on SRAM arrays, which cost a lot in terms of area and power.
		As to key storage, \cite{huang_new_2020} is slightly better than the proposed methods. 
		\cite{zou_security_2020}'s key storage is 1.43 times that of ours because of its larger number of MUX/DEMUX.

		\begin{table}
			\centering
			\caption{Area and power overheads of the proposed methods}
			\vspace{-7pt}
			\label{tab:overhead_area_power}
			\scalebox{0.9}{
				\begin{tabular}{|c|c|c|c|c|c|}
					\hline
					\multicolumn{2}{|c|}{}     &LeNet    &AlexNet  &VGG16    \\ \hline
					\multirow{2}{*}{Area}  &m1 &0.0048\% &0.1108\% &0.0197\% \\ \cline{2-5}
					                       &m2 &0.0011\% &0.0246\% &0.0044\% \\ \hline
					\multirow{2}{*}{Power} &m1 &0.0341\% &0.0975\% &0.0145\% \\ \cline{2-5}
					                       &m2 &0.0096\% &0.0265\% &0.0039\% \\ \hline
				\end{tabular}
			}
			\vspace{-16pt}
		\end{table}

		\begin{table}
			\centering
			\caption{Normalized results of the comparison of area, power, and key storage overheads for the proposed methods and the related work}
			\vspace{-7pt}
			\label{tab:overhead_comparison_normalized}
			\scalebox{0.77}{
				\begin{tabular}{|c|c|c|c|c|c|c|c|c|c|}
					\hline
					                               &\multicolumn{3}{|c|}{Mapping scheme 1} &\multicolumn{3}{|c|}{Mapping scheme 2} \\ \hline
					                               &Area     &Power    &Key Storage        &Area     &Power    &Key Storage \\ \hline
					Our                            &1X       &1X       &1X                 &1X       &1X       &1X          \\ \hline
					date20\cite{zou_security_2020} &--       &--       &--                 &64.80X   &64.80X   &1.43X       \\ \hline
					asp21\cite{wang_low_2021}      &18.00X   &18.00X   &1.02X              &43.20X   &43.20X   &1.02X       \\ \hline
					sram20\cite{huang_new_2020}    &6417.29X &408.19X  &0.96X              &3439.05X &979.65X  &0.96X       \\ \hline
				\end{tabular}
			}
			\vspace{-16pt}
		\end{table}

	Overall, the proposed methods outperform the related work when considering both the effectiveness and hardware overhead at the same time.
	
\vspace{-6pt}
\section{Conclusion}
\IEEEpubidadjcol
    \vspace{-6pt}
	Memristor computing systems are vulnerable to theft attacks intending to steal the NN weights stored in the memristor crossbars. 
	To thwart the attack, this paper proposed an effective countermeasure based on selectively encoding some columns of weights.
	Furthermore, we implemented the designed protection methods in memristor computing systems and added additional techniques to increase their security.
	The experimental results show that the proposed method achieves effectiveness comparable to the best of prior work and imposes the lowest hardware overhead on the memristor computing systems.

\vspace{-8pt}
\bibliographystyle{ieeetr}
\bibliography{references}

\end{document}